# Calibration offset estimation in mobile hearing tests via categorical loudness scaling

Chen Xu[1], and Birger Kollmeier[1]

[1] Medizinische Physik and Cluster of Excellence Hearing4all, Universität Oldenburg, D-26111 Oldenburg, Germany

Corresponding author:
Chen Xu
chen.xu@uni-oldenburg.de
Department of Medical Physics and Acoustics, Faculty VI
University of Oldenburg, 26111, Oldenburg, Germany

**Calibration offset estimation in mobile hearing tests via categorical loudness scaling**

**Objective:**
To enable reliable smartphone-based hearing assessments by developing methods to estimate device calibration offsets using categorical loudness scaling (CLS).

**Design:**
Calibration offsets were simulated from a Gaussian distribution. Two prediction models—a Bayesian regression model and a nearest neighbor model—were trained on CLS-derived parameters and data from the Oldenburg Hearing Health Repository (OHHR). CLS was chosen because it provides level-independent measures (e.g., dynamic range) that remain robust despite calibration errors.

**Study Sample:**
The dataset comprised CLS results from $N = 847$ participants with a mean age of 70.0 years (SD = 8.7), including 556 male and 291 female listeners with diverse hearing profiles.

**Results:**
The Bayesian regression model achieved correlations of up to 0.81 between estimated and true calibration offsets, enabling accurate individual-level correction. Compared to threshold-based approaches, calibration uncertainty was reduced by factors between 0.41 and 0.79, demonstrating greater robustness in uncontrolled environments.

**Conclusions:**
CLS-based models can effectively compensate for missing calibration in mobile hearing assessments. This approach provides a practical alternative to threshold-based methods, supporting the use of smartphone-based tests outside laboratory settings and expanding access to reliable hearing healthcare in everyday and resource-limited contexts.

**Introduction**
Smartphone-based listening tests offer convenient, accessible, and cost-effective solutions for auditory assessment (Xu et al., 2024a; 2024b; 2024c; 2024d; 2025). However, unlike conventional clinical assessments where equipment is carefully calibrated, mobile devices are typically not calibrated. This lack of calibration can compromise the precision of stimulus presentation, potentially leading to inaccurate and unreliable test results.

Hearing threshold measurements, such as those obtained from audiograms, are known to be highly sensitive to device calibration. In contrast, certain supra-threshold listening tests—such as speech-in-noise tests (e.g., Zokoll et al., 2012; Vlaming et al., 2014; Almufarrij et al., 2023)—tend to be, at least partially, robust to uncalibrated playback devices because they primarily rely on relative parameters (e.g., level differences).

When difference measures are used, the results are largely independent of absolute level, reducing their dependence on calibrated playback conditions. This principle has been successfully exploited in speech recognition threshold (SRT) tests and may also apply to other supra-threshold differential measures, such as dynamic range or slope values derived from categorical loudness scaling (CLS). If several of these parameters can be reliably assessed despite calibration offsets—and still characterize the individual's hearing profile with reasonable precision—they may, in turn, be used to estimate the calibration offset itself. In this study, we focus on using a supra-threshold listening test, namely categorical loudness scaling (CLS), to estimate calibration offsets. This study investigates the feasibility of this approach.

The CLS test is designed to assess an individual's perception of loudness. Its adaptive implementation, known as adaptive categorical loudness scaling (ACALOS), is described in detail by Brand and Hohmann (2022). In the ACALOS procedure, participants rate the perceived loudness of sounds using an 11-point scale that includes 7 labeled categories (i.e., "not heard," very soft "soft," "medium," loud, very loud, "too loud") and 4 unlabeled intermediate categories. The primary outcome of ACALOS is the loudness growth function, which characterizes the relationship between sound level and perceived loudness. This function is typically described by six parameters:

- mlow: the slope of the function at low levels,
- mhigh: the slope at high levels,
- Lcut: the level at the transition point between the two slopes,
- HTL (L2.5): the hearing threshold level, defined as the level corresponding to 2.5 categorical units (CU),
- MLL (L25): the medium loudness level, at 25 CU, and
- UCL (L50): the uncomfortable loudness level, at 50 CU.

Moreover, the dynamic range (DR) is defined as the difference in sound level between the UCL and HTL. When using uncalibrated mobile devices, parameters such as HTL, MLL, UCL, and Lcut are assumed to be influenced by calibration offsets. These parameters are therefore referred to as level-dependent or calibration-variant in the context of the CLS test. In contrast, the parameters mlow, mhigh, and DR are considered unaffected by calibration, as they represent relative quantities—either slopes or level differences—rather than absolute sound levels. As such, they are classified as level-independent or calibration-invariant parameters in the CLS test.

To evaluate whether supra-threshold CLS parameters can be used to estimate playback calibration offsets, we first introduce simulated calibration offsets, drawn from Gaussian distributions, to a calibrated clinical dataset—the Oldenburg Hearing Health Repository (OHHR; see Jafri et al., 2025). We then use the level-independent parameters of the CLS test (i.e., mlow, mhigh, and DR) to estimate the calibration offset through two modeling approaches: a regression-based model and a nearest-neighbor model. Finally, the estimated calibration offsets are compared against the true (simulated) offsets. The overarching aim is to evaluate whether a supra-threshold CLS test, performed without prior calibration, can provide a sufficiently accurate estimate of the calibration offset—within approximately 5 dB—by leveraging a large, calibrated auditory dataset (OHHR). The following three research questions (RQs) will be addressed:

1. **RQ1:** Is it feasible to estimate the calibration offset using the uncalibrated CLS test?
2. **RQ2:** What factors (e.g., estimation models, number of frequencies, auditory profiles) influence the calibration error?
3. **RQ3:** To what extent can the calibration offset uncertainty be reduced by the proposed estimation procedure?

**Methods**
*Oldenburg Hearing Health Repository*
The Oldenburg Hearing Health Repository (OHHR) was used in this study, comprising data from N = 847 participants with a mean age of 70.0 years (SD = 8.7). The dataset included 556 male and 291 female participants. All participants had pure-tone averages (PTA4; average hearing thresholds at 0.5, 1, 2, and 4 kHz) greater than 20 dB HL. Hearing profiles were categorized according to Bisgaard et al. (2010), with the following distribution: 275 (N3), 221 (N2), 121 (S1), 95 (S2), 80 (N4), and 55 (S3) participants.

All participants completed the standard ACALOS procedure (ISO 16832, 2006) using Sennheiser HDA200 headphones (Brand & Hohmann, 2002; Oetting et al., 2014). The experimental setup was fully calibrated. Narrowband noises (2-second duration) centered at 1500 and 4000 Hz were presented adaptively to each ear. For further details on data collection, see Jafri et al. (2025).

*Simulation of Calibration Offsets*
To simulate measurements from uncalibrated mobile devices, we generated calibration offsets from normal distributions with mean values of 5 and 10 dB, each with standard deviations (SD) of 5 and 10 dB. These offsets were then added to the calibrated dataset described above. Please note that the calibration offsets were applied only to the four level-dependent parameters (i.e., Lcut, HTL, MLL, and UCL), while the three level-independent parameters (i.e., mlow, mhigh, and dynamic range) remained unaffected.

*Estimation of Calibration Offsets*
a) regression-based model

Calibration offset estimation using a regression-based approach was conducted as follows. First, two Bayesian regression models were trained to predict L25 (the medium loudness level) in the calibrated dataset using six level-independent parameters: mlow, mhigh, and dynamic range (DR) at 1500 Hz and 4000 Hz. Figure 1 illustrates the comparisons between the empirical and model-predicted L25 values.

Second, the trained models were applied to the uncalibrated datasets by inputting mlow, mhigh, and DR to predict L25. Third, calibration offsets were estimated by computing the difference between the predicted L25 (averaged across the two frequencies) and the observed L25 in the uncalibrated dataset. Finally, the estimated calibration offsets were compared to the true calibration offsets, as shown in Figure 3.

The Bayesian regression models were fitted using the "brms" package in R (Bürkner, 2017; Carpenter et al., 2017), and scatter plots were generated using the "ggpubr" package (Kassambara, 2023).

b) nearest-neighbor model
For comparison, a nearest-neighbor approach was implemented as follows. First, we matched the six level-independent parameters (i.e., mlow, mhigh, and DR at two frequencies) from the uncalibrated dataset to those in the calibrated dataset by minimizing the root mean squared error (RMSE). This step identified the "nearest neighbor" based solely on the level-independent parameters. Second, we estimated the calibration offset as the difference in L25 between the uncalibrated measurement and its nearest neighbor in the calibrated dataset. Finally, we compared the estimated offset to the true offset (see Figure 4 for results).

*Statistical Analysis*
To assess the agreement between the estimated and true calibration offsets, we employed the Pearson correlation coefficient (PCC). Corresponding p-values were reported to indicate the statistical significance of the correlations. To address research question 3 (RQ3), we calculated calibration uncertainty reduction factors, defined as the standard deviation of the estimated calibration offsets divided by the corresponding "true" calibration offsets (known a priori; see Table 3 for results).

**Results**
**A) Self-consistency:** As a first step, we examined the self-consistency of the underlying OHHR database, specifically the extent to which calibration-offset–dependent parameters could be predicted from offset-independent parameters.

*L25 prediction using the Bayesian regression model*

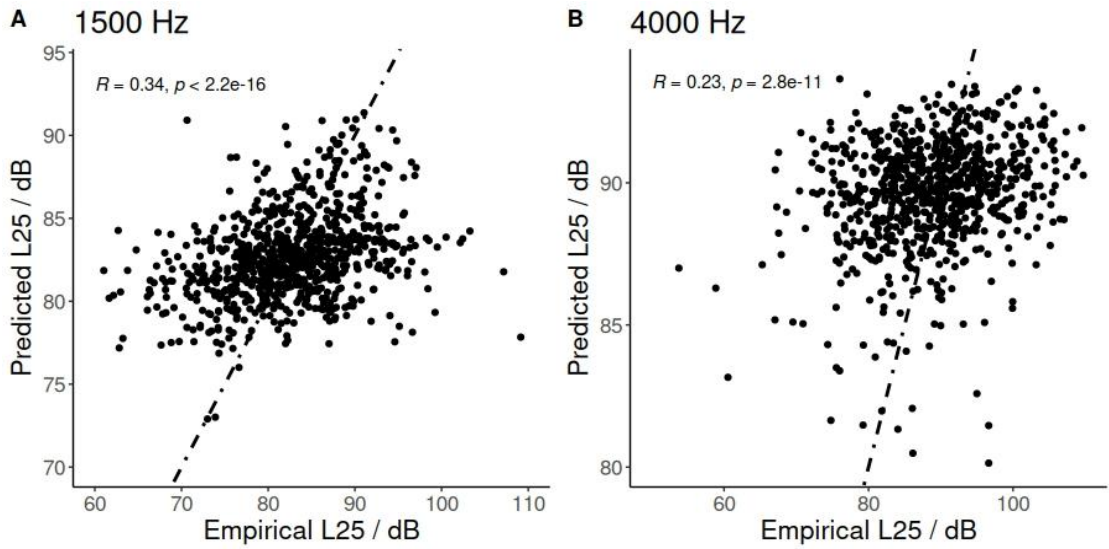

Figure 1. Predicted L25 values (medium loudness level, in dB) from the Bayesian regression models using mlow, mhigh, and dynamic range (DR) at two frequencies (1500 Hz and 4000 Hz) as input values plotted against the "true" empirical L25 values for N = 847 participants. Panels A and B show results for 1500 Hz and 4000 Hz, respectively. The dot-dashed line represents the identity line (slope = 1). The correlation coefficient (R) and p-value are indicated in the upper-left corner of each panel.

Figure 1 shows the simulated predicted L25 values plotted against the empirical L25 values using the Bayesian regression model. At both frequencies, the predicted values were significantly positively correlated with the empirical values ($p < 0.05$). The correlation coefficients were 0.32 at 1500 Hz and 0.24 at 4000 Hz, indicating relatively poor predictability of the calibration-dependent supra-threshold parameter L25 from calibration-independent parameters at these frequencies.

*L2.5 prediction using the Bayesian regression model*

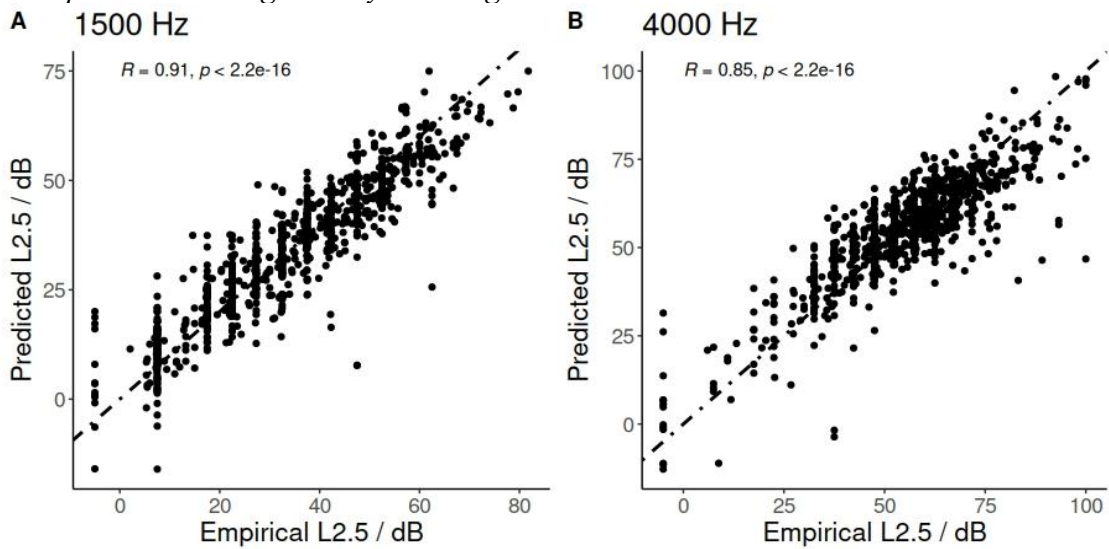

Figure 2. Predicted versus empirical L2.5 values (estimated hearing threshold levels in dB) based on a Bayesian regression model using six input parameters (slopes mlow, mhigh, and dynamic range at two frequencies) for N = 847 participants. Panel A shows results at 1500 Hz, and Panel B at 4000 Hz. Please see Figure 1 for details on the dot-dashed identity line and the reported correlation coefficients.

Figure 2 illustrates the self-consistency and accuracy of L2.5 values predicted by the regression model based on supra-threshold CLS parameters, compared to the original measured L2.5. Overall, the predicted values were significantly positively correlated with the measured L2.5 values at both frequencies ($p < 0.05$), with correlation coefficients of 0.91 and 0.85, indicating strong correlations.

Table 1. Median absolute error (MAE, in dB) between measured and predicted L2.5 values at 1500 Hz and 4000 Hz, reported for all participants (N = 847) and as a function of Bisgaard profiles (see Bisgaard et al., 2010, for details).

|         | Overall | N2   | N3   | N4   | S1   | S2   | S3   |
|---------|---------|------|------|------|------|------|------|
| 1500 Hz | 4.37    | 4.61 | 4.17 | 6.01 | 3.35 | 3.98 | 5.74 |
| 4000 Hz | 5.56    | 6.75 | 5.19 | 5.04 | 4.68 | 5.67 | 7.49 |

Table 1 presents the median absolute errors (MAEs) between measured and predicted L2.5 values obtained using the regression model. On average, MAEs were approximately 5 dB for both frequencies. At 1500 Hz, the highest MAE was observed in the N4 group, and the lowest in the S1 group. At 4000 Hz, the S3 group showed the highest MAE, while S1 again showed the lowest. Overall, MAE values at 1500 Hz were generally lower than those at 4000 Hz. In addition, the errors generally increased with the degree of hearing loss, suggesting reduced prediction accuracy for individuals with more severe impairments.

**B) Calibration offset estimation**
In a second step, we used the calibration-offset–independent parameters from both measurement frequencies to estimate the calibration offset, which was assumed to be identical across frequencies.

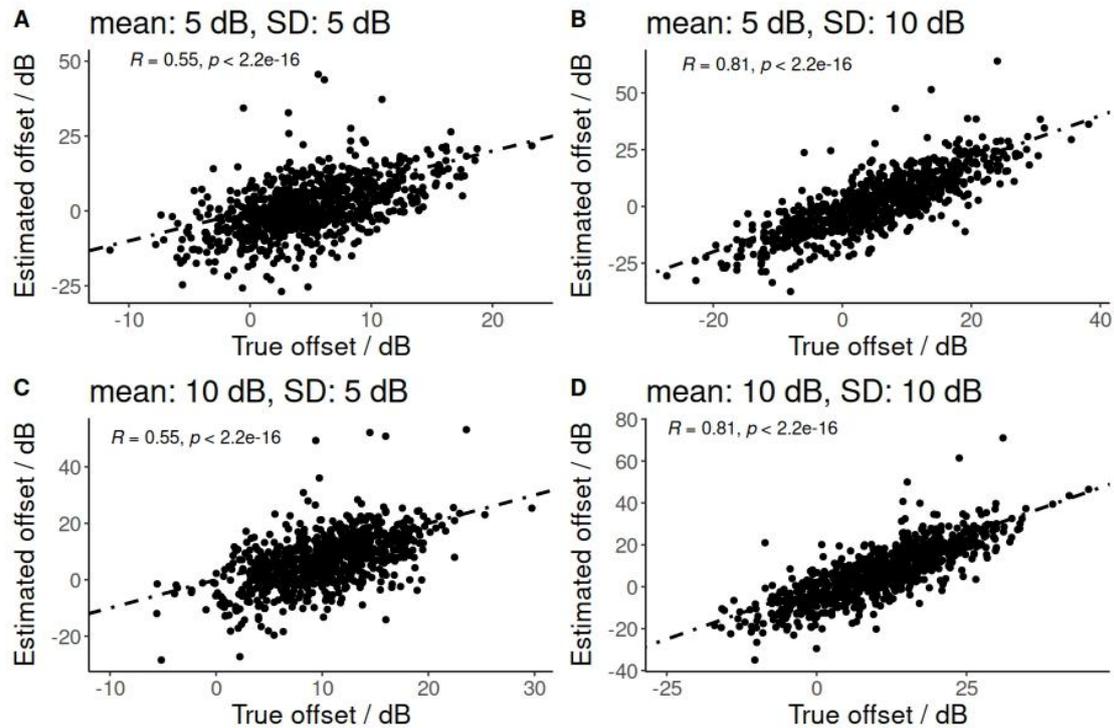

Figure 3. Estimated calibration offset (in dB) using the regression-based model plotted against the "true" offset on N = 847 participants. True calibration offsets are simulated by sampling from Gaussian distributions with the following parameters: (A) mean = 5 dB, SD = 5 dB; (B) mean = 5 dB, SD = 10 dB; (C) mean = 10 dB, SD = 5 dB; and (D) mean = 10 dB, SD = 10 dB. The dot-dashed line represents the identity line (slope = 1).

Estimated calibration offsets derived from the Bayesian regression-based models are shown as a function of the true offsets in Figure 3. Overall, all estimated offsets were significantly positively correlated with the true offsets (p < 0.05). Specifically, when the true offsets were drawn from distributions with a mean of 5 or 10 dB and a standard deviation (SD) of 5 dB (Panels A and C), the correlation coefficient was 0.55, indicating a moderate predictability of the calibration offset. In contrast, when the SD was increased to 10 dB (Panels B and D), the correlation coefficient rose to 0.81, indicating a very strong predictability of the offset. Moreover, the mean values of the "true" calibration offsets appeared to have no effect on the estimated offsets, indicating the absence of systematic bias. The shifts were perfectly aligned.

The results shown in Figures 2 and 3 and Table 1 demonstrate that calibration offset estimation errors arise from two main sources: (1) statistical limitations of the regression model, reflected in the limited self-consistency of the underlying data—particularly in listeners with more severe or atypical audiograms—and (2) the limited predictive power of calibration-offset-independent parameters when used to estimate calibration-offset-dependent quantities such as the "true" calibration offset. Nevertheless, combining estimates from two independent frequencies yields a measurable reduction in the common offset estimation (see below). Please note that the distribution of the "true" calibration offsets only has a limited influence.

Table 2 summarizes the median absolute errors (MAEs) between the estimated and "true" calibration offsets (**mean = 10 dB, SD = 10 dB**) for all 847 participants, as well as for subgroups defined by Bisgaard profiles. The regression-based model generally yielded MAEs below 5 dB, whereas the nearest-neighbor model showed MAEs ranging from 5 to 10 dB, suggesting that the regression-based model may be more suitable for practical applications. Among Bisgaard subgroups, the regression-based model produced the lowest MAE in the N4 group and the highest in the N2 group. For the nearest-neighbor model, the S1 profile showed the lowest MAE, while the S3 profile had the highest.

Moreover, Table 3 presents the calibration uncertainty reduction factors resulting from the estimation process, defined as the ratio between the standard deviation of estimated calibration offsets and that of the true calibration offsets. Across all parameter sets, the factors were consistently below 1, indicating that our proposed approach reduced uncertainty. Furthermore, the reduction factors increased with larger standard deviations of the true offsets, suggesting a dependence on the variability of the true calibration offset.

Table 2. Median absolute error (MAE, in dB) between estimated and "true" calibration offsets **(mean = 10 dB, SD = 10 dB)** across all participants (N = 847) and within subgroups defined by Bisgaard profiles (see Bisgaard et al., 2010, for profile definitions).

|  | Overall | N2 | N3 | N4 | S1 | S2 | S3 |
|---|---|---|---|---|---|---|---|
| regression-based | 5.06 | 6.01 | 5.16 | 3.69 | 5.92 | 4.33 | 4.45 |
| nearest-neighbor | 6.59 | 6.57 | 6.33 | 7.65 | 5.89 | 6.66 | 9.64 |

Table 3. Calibration uncertainty reduction factors for two prediction models. The factor is defined as the ratio of the standard deviation (SD) of estimated calibration offsets to that of the true calibration offsets. Results are shown for four simulated offset distributions: (5, 5) = mean 5 dB, SD 5 dB; (5, 10) = mean 5 dB, SD 10 dB; (10, 5) = mean 10 dB, SD 5 dB; (10, 10) = mean 10 dB, SD 10 dB.

|  | (5,5) | (5,10) | (10,5) | (10,10) |
|---|---|---|---|---|
| regression-based | 0.55 | 0.78 | 0.55 | 0.79 |
| nearest-neighbor | 0.41 | 0.67 | 0.41 | 0.68 |

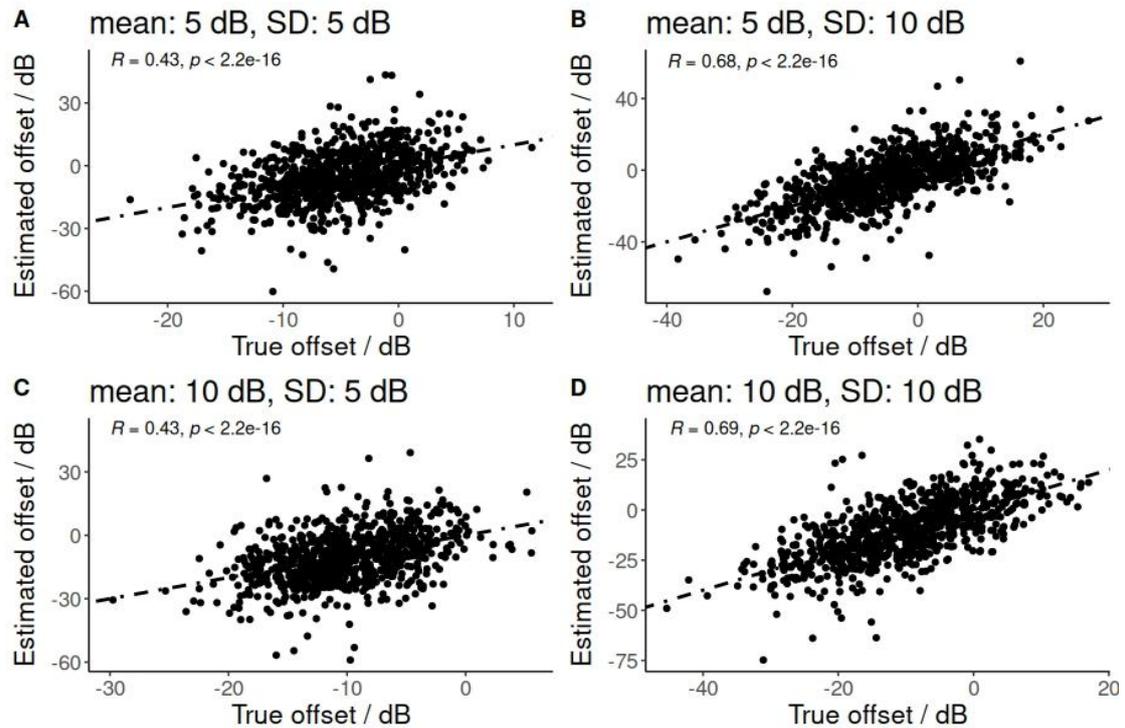

Figure 4. Scatter plots of the estimated calibration offsets (in dB) obtained using the nearest-neighbor model plotted against the true offsets on N = 847 participants. Refer to Figure 3 for details on the sub-panels, the dot-dashed identity line, and the statistical annotations.

Figure 4 shows the estimated calibration offsets derived from the nearest neighbor approach plotted against the true offsets. In all sub-panels, the estimated offsets were significantly positively correlated with the true offsets (p < 0.05). As with the regression model (cf. Figure 3), the correlation coefficients increased when the standard deviation (SD) of the true offset rose from 5 dB to 10 dB. Compared to Figure 3, however, the Bayesian regression model yields higher correlation coefficients across all conditions, suggesting that its estimated offsets are more closely aligned with the true offsets than those from the nearest neighbor approach (See also Table 2, which presents the median absolute errors between the estimated calibration offsets and the "true" calibration offset (mean = 10 dB, SD = 10 dB). This indicates that the Bayesian regression model may provide more accurate predictions of calibration offsets.

**Discussion**

In this study, we investigated the estimation of calibration offsets from supra-threshold uncalibrated categorical loudness scaling (CLS) data by employing a large auditory dataset from the Oldenburg Hearing Health Repository as reference. True calibration offsets were simulated by sampling from Gaussian distributions with mean values of 5 or 10 dB and standard deviations of 5 or 10 dB. Two modeling approaches were evaluated: a Bayesian regression model and a nearest-neighbor model. Our findings demonstrate that it is feasible to estimate calibration offsets from CLS data, with the estimated values showing good agreement with the simulated true offsets. Notably, the Bayesian regression model achieved a correlation coefficient of up to 0.81, suggesting a high level of accuracy in offset estimation under certain conditions.

Moreover, the calibration offset reduction factor ranged from 0.41 to 0.79, indicating that our estimation models can substantially reduce the uncertainty of the calibration offset.

Previous studies (Kisić et al., 2022; Scharf et al., 2024) have typically employed either speech-based or bottle-whistle-based procedures to calibrate smartphones. In the speech-based approach, participants adjusted the loudness of pre-recorded speech to match live speech read aloud at 55 dB(A). In the bottle-whistle-based method, participants produced a whistling tone by blowing across the opening of an empty glass bottle and recorded the resulting sound level using a mobile device. Both methods require additional materials (e.g., a bottle, a human reader), which may limit their practicality and scalability. In contrast, the present approach requires no additional resources beyond an uncalibrated categorical loudness scaling (CLS) test—which is typically included in the audiological evaluation for each subject—and, during the model training phase, access to a large auditory database. This makes our method more cost-effective, user friendly, and secure than previously proposed calibration procedures for uncalibrated remote auditory test devices: Participants simply perform the uncalibrated CLS test, and calibration offsets are estimated and corrected automatically. This reduces the burden on users and enhances the feasibility of remote listening tests on mobile devices. Finally, our method is data-driven—that is, calibration estimates are derived from empirical data rather than subjective judgments—thereby minimizing potential bias. In addition, the estimated offsets are individualized, allowing for personalized and adaptive calibration for each user.

Two modeling approaches—the Bayesian regression-based model and the nearest-neighbor model—were compared in this study for estimating calibration offsets. Our results indicate that the Bayesian regression approach outperforms the nearest-neighbor method in terms of prediction accuracy, as reflected by higher correlation coefficients with the true offsets. The superior performance of the Bayesian model may be attributed to its ability to incorporate prior information and handle uncertainty, consistent with previous findings in auditory modeling (e.g., McMillan & Cannon, 2019). Additionally, the Bayesian regression model accounts for covariance between input variables, which may further enhance its predictive capabilities. Based on these findings, we recommend the Bayesian regression model for future applications in calibration offset estimation.

The key operational principle of the present approach is the assumed constancy of the calibration factor across a specific audio frequency range, which is expected to reduce estimation error as the number of independent CLS observations across frequencies increases. In this study, we demonstrate feasibility using only two frequencies (1500 Hz and 4 kHz); however, further reductions in uncertainty are anticipated as the number of frequencies increases. Additionally, the accuracy of the estimated loudness growth function at each frequency improves with a larger number of stimuli at varying presentation levels, providing more level-independent information that can be leveraged for a more precise calibration factor estimate. This, however, comes at the cost of longer measurement times for each subject. Consequently, the well-known trade-off between time and precision must be considered not only for the psychophysical data collected but also for the calibration offset derived from these data.

A key strength of the approach in our study lies in the use of simulated calibration offsets with systematically controlled variability, enabling a direct and fair comparison of model performance under different statistical conditions. However, a primary limitation is that the evaluation relied on simulated data, which may not fully reflect the variability encountered in real-world calibration scenarios. As a next step, we therefore plan to collect data from listening tests conducted with uncalibrated mobile devices and validate the proposed approaches using empirical measurements.

**Limitations and Outlook**

Although the approach presented here, along with its evaluation and characterization of statistical properties, appears promising for real-world applications with remote auditory test devices, several caveats and limitations must be acknowledged:

1. **Simulation-based evaluation:** This study is based solely on simulations. Practical validation of the method in real-world settings will be conducted in future work.

2. **Range of simulated calibration offsets:** The simulated offset distributions cover a limited range, as a detailed investigation of the systematic influence of offset distribution—likely to be minor—was beyond the present scope. Nevertheless, the practical range of calibration offsets (approximately 20–30 dB) is well represented by the assumed "true" offset distributions with standard deviations of 5 dB and 10 dB.

3. **Restricted frequency range:** Only two frequencies were included in the simulations due to limitations of the OHHR database.

4. **Choice of supra-threshold measurement:** ACALOS was selected as the basis for supra-threshold measurements, supported by its use in auditory profiling (Saak et al., 2025; Xu et al., 2024d) and in hearing aid prescription and fitting (Exter et al., 2025). In future work, ACALOS may be replaced by rACALOS (Xu et al., 2024c), which enables more accurate estimation of absolute thresholds and thus potentially yields more precise calibration factors. In principle, other supra-threshold psychophysical procedures could also be employed, provided they yield level-independent parameters that correlate—at least weakly—with the absolute threshold or audiogram. In this regard, CLS represents a strong candidate, as the shape of the loudness growth function is, on average, highly dependent on the absolute threshold.

**Conclusions**

This study demonstrates the feasibility of inferring calibration offsets from uncalibrated CLS tests using level-independent parameters in combination with a large auditory database. Although only a limited range of factors affecting calibration offset estimation accuracy was considered, the method proved relatively robust to the distribution of "true" calibration offsets when CLS data from two frequencies were used. Using the proposed Bayesian regression model, the estimated calibration offsets exhibited strong correlations with the true values and reduced level calibration uncertainty by a factor ranging from 0.55 to 0.79.

These findings suggest that the approach can be effectively applied in mobile hearing assessments, even when using uncalibrated equipment.

**Appendix: Estimation of calibration offsets using a single frequency (4000 Hz)**

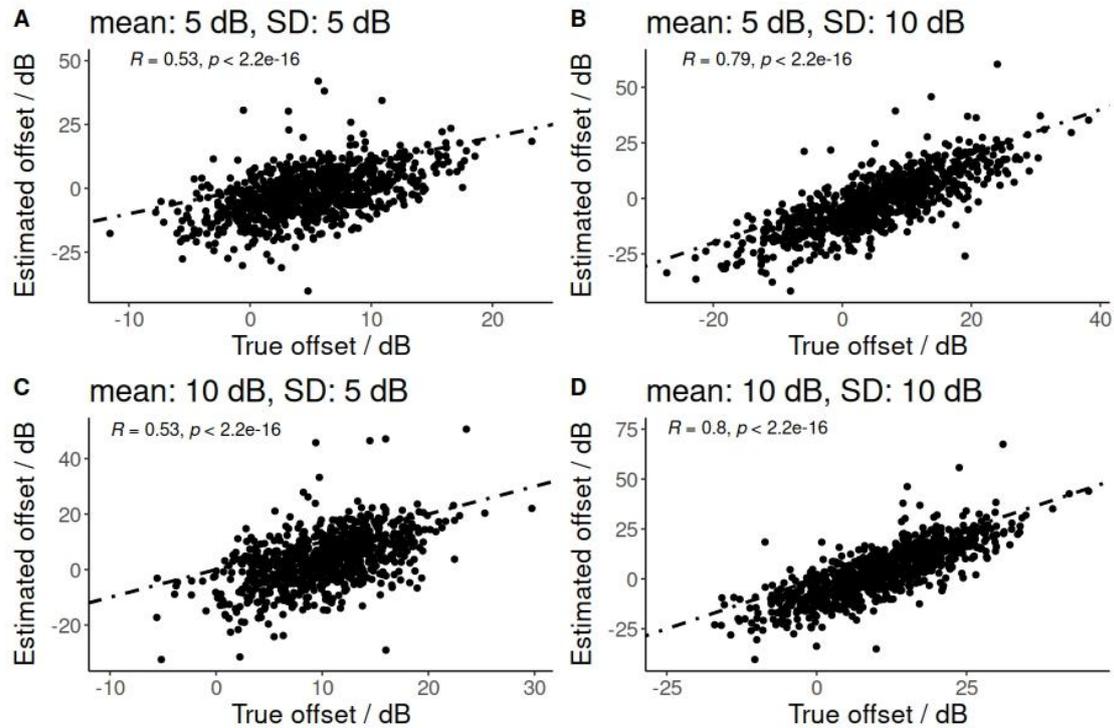

Figure A1. Estimated vs. true calibration offsets using CLS data from a single frequency (4000 Hz) across N = 847 participants. Refer to Figure 3 for details on the sub-panels, dot-dashed identity line, and statistical annotations.

Compared to Figure 3, which used data from two frequencies, Figure A1 presents calibration offset estimates based on CLS data from only one frequency. The correlation coefficients obtained with a single frequency were slightly lower than those derived from two frequencies. The overall median absolute error (MAE) across all participants was 7.42 dB, while the MAEs for the N2, N3, N4, S1, S2, and S3 Bisgaard profiles were 8.79, 7.84, 4.00, 8.31, 7.38, and 4.35 dB, respectively. These values were significantly higher than those obtained using two frequencies ($p < 0.05$; see Table 2). Moreover, a significant estimation bias is evident, as the distribution is asymmetrical with respect to the identity line, whereas no such bias was observed in Figure 3. This indicates that estimating calibration offsets using two frequencies yields more accurate results than using only one, which is expected as incorporating data from two frequencies provides more information and thus improves estimation accuracy. Overall, calibration offset estimation errors were reduced by approximately 2.5 dB when two frequencies were used.

**Acknowledgments**
This work was funded by the Deutsche Forschungsgemeinschaft (DFG, German Research Foundation) under Germany's Excellence Strategy – EXC 2177/1 - Project ID 390895286.

**Disclosure statement**

No potential conflict of interest was reported by the author(s).